\documentclass{emulateapj}
\usepackage{graphicx}
\usepackage{epstopdf}
\def\ie{{\it i.e., }}
\def\be{\begin{equation}}
\def\ee{\end{equation}}
\def\bea{\begin{eqnarray}}
\def\eea{\end{eqnarray}}
\newcommand{\eg}{{\it e.g., }} 

\journalinfo{Astrophysical Journal, in press}
\slugcomment{}

\shorttitle{SED fitting with MCMC}
\shortauthors{Acquaviva et al.}

\begin{document}

\title{SED fitting with Markov Chain Monte Carlo: \\
Methodology and Application to $z$=3.1 Ly$\alpha$-Emitting Galaxies}

\author{Viviana Acquaviva\altaffilmark{1}, Eric Gawiser\altaffilmark{1}, Lucia Guaita\altaffilmark{2,3}}

\altaffiltext{1}{Department of Physics and Astronomy, Rutgers, The State University of New Jersey, Piscataway, NJ 08854}
\altaffiltext{2}{Departimento de Astronomia y Astrofisica, Universidad Catolica de Chile,
    Santiago, Chile}
     \altaffiltext{3}{Institutionen f{\"o}r Astronomi, Stockholms Universitet, SE-106 91 Stockholm, Sweden}
    
\begin{abstract}
We present GalMC, a MCMC algorithm designed to fit the spectral energy distributions (SED) of galaxies to infer physical properties such as age, stellar mass, dust reddening, metallicity, redshift, and star formation rate. We describe the features of the code 
and the extensive tests conducted to ensure that our procedure leads to unbiased parameter estimation and accurate evaluation of uncertainties. We compare its performance to grid-based algorithms, showing that the efficiency in CPU time is $\sim$ 100 times better for MCMC for a three dimensional parameter space and increasing with the number of dimensions. 
We use GalMC to fit the stacked SEDs of two samples of Lyman Alpha Emitters (LAEs) at redshift $z$=3.1. 
Our fit reveals that the typical LAE detected in the IRAC 3.6 $\mu$m band has age = 0.67 [0.37 -1.81] Gyr and stellar mass = 3.2 [2.5 - 4.2] $\times10^9$ M$_\odot$, while the typical LAE not detected at 3.6 $\mu$m has age = 0.06 [0.01-0.2] Gyr and stellar mass = 2 [1.1 - 3.4] $\times 10^8$ M$_\odot$.
The SEDs of both stacks are consistent with the absence of dust. The data do not significantly prefer exponential with respect to constant star formation history. 
The stellar populations of these two samples are consistent with the previous study by Lai et al, with some differences due to the improved modeling of the stellar populations. A constraint on the metallicity of $z$=3.1 LAEs from broad-band photometry, requiring $Z < Z_\odot$ at 95\% confidence, is found here for the first time.
\end{abstract}

\section{Introduction}

A galaxy's Spectral Energy Distribution (SED) contains information about its stellar population age, mass, star formation rate (SFR), dust content, and metallicity.
Different physical processes leave their imprint in different parts of the spectrum; the wider the wavelength coverage, the more robust the interpretation of the various features of the SED is in terms of galaxy properties.
The rest-frame ultraviolet (UV), optical, and infrared (IR) parts of the spectrum offer the combination of depth and angular resolution needed to obtain individual photometry for all galaxies detected in a particular band, which is not generally possible in the far-infrared through sub-millimeter wavelengths where re-emission by dust dominates the luminosity.
Spectroscopic information is generally required to determine metallicity and further probes of chemical enrichment, but spectroscopy of unbiased samples of dim galaxies is very difficult to obtain.
This leaves photometric UV-through-IR SEDs as the most readily available probe of galaxy properties.

SED fitting is the procedure of comparing models to the observed galaxy SED (for reviews, see \citealt{2009NewAR..53...50G,2010Ap&SS.tmp..257W} and references therein). Since the physical properties of the models are known, those of the data can be derived by maximizing the resemblance between data and models. The success and reliability of this method depend on the quality of the available template spectra and on the robustness of the fitting algorithm. 
The ingredients of the spectral templates typically are libraries of stellar spectra and sets of evolutionary tracks, allowing one to compute the initial spectrum of a collection of stars of different mass and to follow its evolution with time. Measuring and calibrating these quantities is not an easy process, and the available models cannot uniformly cover stellar populations of all ages, masses, or chemical content. The large variance intrinsic in the spectral templates is reflected in the large number of available stellar population synthesis (SPS) codes, including but not limited to PEGASE \citep{1997A&A...326..950F}, Maraston \citep{1998MNRAS.300..872M,2005MNRAS.362..799M}, GRASIL \citep{1998ApJ...509..103S}, Starburst99 \citep{1999ApJS..123....3L,2005ApJ...621..695V}, GALAXEV (Bruzual and Charlot 2003, Charlot and Bruzual 2010), and GALEV \citep{2009MNRAS.396..462K}. The scatter among the predictions of different SPS models is in itself an important source of systematic uncertainty (see \eg \citealt{2007ApJ...657L...5K,2010ApJ...712..833C} and our discussion in Sec. \ref{sec:results}). 

Often, the comparison between model and data is made through a $\chi^2$ minimization, which provides us with a best-fit model. If the probability distribution of the parameters is close to a Gaussian (\eg \citealt{2010arXiv1001.4635L}), the best fit is a good estimate of the expectation value for each parameter, and the corresponding uncertainties can be evaluated from relative likelihoods assuming a Gaussian profile.
This approach can, however, be very dangerous in astronomy. In fact, because of the unprecedented volume of available data, we can now hope to explore new aspects of the physics of galaxies, and this exciting perspective comes at the price of not having prior knowledge of the probability distribution we set out to investigate. Furthermore, the high degree of correlation between astrophysical processes makes probability distributions highly non-Gaussian. In a multi-dimensional space with heavy parameter degeneracies, the ``best fit" spectrum is an overly aggressive compression of the information available in the SED, and the assumption of a Gaussian likelihood is unjustified. As a result, it is necessary to switch to more sophisticated and robust statistical analysis techniques, such as Markov Chain Monte Carlo (MCMC) algorithms.  
While the use of simple $\chi^2$ minimization techniques is still very common, MCMC is becoming a popular statistical method within the SED fitting community (\eg \citealt{2003MNRAS.343.1145P,2006MNRAS.369..939S,2007A&A...471...71N,2010arXiv1009.0007N,2010ApJ...712..833C, 2011arXiv1103.3269S, pirzkal}, Johnson et al 2011 in prep); however, none of these algorithms is publicly available yet. Our MCMC code, GalMC, is available at \\ 
\small http://www.physics.rutgers.edu/$\sim$vacquaviva/web/GalMC.html.
\normalsize

This paper is organized as follows. In Sec. \ref{sec:MCMC} we present the general features of MCMC algorithms and describe in the detail GalMC, our MCMC code for SED fitting. Sec. \ref{sec:test} illustrates the extensive tests we performed on Mock catalogs to ensure that the algorithm works correctly and that our inferred parameter values and uncertainties are trustworthy. In Sec. \ref{sec:results} we present the results we obtain for two samples of Lyman Alpha Emitters (LAE) at redshift $z$ = 3.1, for various assumptions of star formation histories (SFH) and stellar populations synthesis (SPS) models; we also compare our results to those obtained for the same samples, but with a different technique, by Lai et al 2008 (hereafter L08). We summarize our findings in Sec. \ref{sec:conc}.

\section{The MCMC algorithm for SED fitting}
\label{sec:MCMC}

The process of parameter fitting involves two essential steps. First, we want to know what is the most representative value, or estimate, of the true value of each parameter included in our fit. Second, we want to evaluate the \textit{uncertainties}, \ie how likely it is that the true value lies in a given interval in the vicinity of that estimate. In order to accomplish this task, one needs to determine the probability density function of the parameters, $p({\bf x})$, which can be used to compute the expectation value of any function of interest as
\be
\label{eq:prob}
\langle f({\bf{x}}) \rangle = \int dx_1... dx_n f({\bf{x}}) p({\bf{x}}) .
\ee
In many cases where the probability density function is complicated and the dimensionality $n$ of the parameter space is high, finding alternate ways of evaluating the above integral is required. This can achieved by informed \textit{sampling} of the parameter space. A sample is a 
vector of coordinates in parameter space; 
the objective of the sampling process 
is to obtain a set of $R$ {\it independent} samples $r_i$ whose distribution is identical to that of the probability density function $p({\bf x})$. In practice, this can be achieved if the density of sampled points is proportional to $p$({\bf x}). Once this collection of samples is obtained, it will by definition satisfy the  property 
\be
\langle f({\bf x}) \rangle = \int dx_1... dx_n f({\bf x}) p({\bf x}) \simeq \frac{1}{R} \sum_{i=1}^R f(r_i) 
\label{eq:fund}
\ee
for any function $f$, and performing integrals of the form of Eq. (\ref{eq:prob}) becomes a trivial matter.   Often, we will be interested in recovering the {\it posterior} probability distribution of parameters. According to Bayesian statistics, which we adopt, this depends on the product of the likelihood of observing the data, if that model were the truth, and on our beliefs on the distribution of parameters, or priors. 
 
A flexible method allowing one to create samples with minimal assumptions on the underlying probability distribution of the parameters with respect to the data is to use Monte Carlo Markov Chains (see \eg \citealt{MacKay, neal93, Lewis:2002ah, 2003ApJS..148..195V, Hajian:2006mt,2010bmic.book.....H}).
Monte Carlo methods are computational techniques that make use of random numbers. In this context, a random number is involved in the process of moving from one point in parameter space to the next one. This succession of steps is called a chain. Because the transition to the next step only depends on the current point, having no memory of the previous points, the chain is called a {\it Markov chain}. 

How can we obtain the desired independent samples from a Markov chain? The chain needs to be built satisfying certain rules that restrict the  {\it transition probability} $T({\bf x},{\bf y})$, the probability of moving from a point $\bf{x}$ to the point $\bf{y}$. Two things are required (\eg \citealt{MacKay}):
\begin{enumerate}
\item The desired probability distribution $p$ is an invariant distribution of the chain:
\be
p({\bf x}) = \int d {\bf y} \, T({\bf x},{\bf y}) \,p({\bf y}).
\ee
\item The chain is {\it ergodic}, \ie  the probability distribution of the chain tends to an invariant distribution of the chain, no matter what the initial conditions are.
\end{enumerate}
If these conditions are satisfied, after an initial period where the walk of the chain depends on the initial conditions (the {\it burn-in} phase), the chain will start sampling from the desired probability distribution (the posterior). However, subsequent steps of the chain are strongly correlated with each other. It is necessary to select points which are distant enough from each other ({\it thin} the chain) in order to achieve the independent samples needed in Eq. \ref{eq:fund}.

There are two main tasks associated with the creation of chain. The first is how to explore the parameter space, namely, how to go efficiently 
from one point to the next one and how to decide whether or not the step just taken will become part of the chain. The latter process generally involves evaluating how similar the model corresponding to a point in parameter space is to the data one wants to fit, based on the model's likelihood given the observed data, and on the chosen priors.
The second task is to compute the likelihood of a given model at each point visited by the chain; we do so by using stellar population synthesis models to predict the observable quantities as a function of the input parameters. These two processes are described in detail below.
 
 \subsection{Description and features of sampling algorithm}
 The structure of our code, {\it GalMC}, is deeply indebted to the publicly available algorithm CosmoMC \citep{Lewis:2002ah}. We use a Metropolis sampling algorithm \citep{Hastings1970}, where a trial step ${\bf x_0} + \Delta{\bf x}$ from the current position in parameter space ${\bf x_0}$ is accepted with probability
 \be
P_{\rm acc} = min\left(1, \frac{p({\bf x_0}+\Delta{\bf x})}{p({\bf x_0})}\right)
\ee
where $p({\bf x_0})$ is the posterior probability evaluated at ${\bf x_0}$.

The trial step ${\bf x_0} + \Delta{\bf x}$ is a random deviate drawn from a distribution, $q({\bf x})$, known as the proposal density. In principle, Metropolis sampling converges for a wide variety of proposal densities; however, in practice, performance is greatly improved upon judicious selection of this distribution. An important indicator of the performance of the sampling algorithm is the {\it acceptance rate}, the ratio between the number of accepted steps and the number of trial steps taken; the rule of thumb for optimal acceptance rate is $\simeq$ 25$\%$ \citep{Roberts97}.   If no previous information on the covariance of parameters (for example, from previous MCMC runs) is available, the proposal density is chosen according to the method of \cite{cosmomc_notes}. A random orthonormal basis in parameter space is chosen; the normalization to orthonormal parameters relies on an initial guess on the width of the target posterior distribution in each direction. Starting from the first direction of the basis vectors, a step of distance r in that direction is drawn from a proposal density consisting of the sum of a Gaussian and an exponential:
\be
q(r) = g \left(\frac{r}{s}\right) e^{(-r/s)^2} + (1-g)e^{-r};
\ee
Combining an exponential tail to a Gaussian proposal density is good practice when target probability distributions are not expected to be Gaussian. The factor $g$ identifies the degree of mixing; we chose the value of $g =2/3$ suggested in \cite{cosmomc_notes} and found nearly optimal acceptance rate, as discussed later in the text. The quantity $s$ regulates the size of a typical trial step with respect to the (estimated) width of the posterior in that direction, and we choose the value $s = 2.4$ as suggested by \cite{Gelman95} and \cite{2005MNRAS.356..925D}. 
We repeat this procedure taking subsequent trial steps in each direction of the basis vectors, then generate a new random basis, and iterate this process.

More efficient sampling is achieved if the covariances between the parameters are known.  When a covariance matrix is available, eigenvectors of this matrix are used to select the basis of orthonormal parameters. 
We introduce the possibility of using an adaptive covariance matrix. This feature is useful in particular for parameter spaces where the shape of the target distribution depends strongly on the position in parameter space, so that previously obtained covariance matrices might not be a good representation of the covariance in the present case. When this option is selected, the sampling process is stopped at intervals that can be chosen by the user, a new covariance matrix is computed on the desired fraction of the chain (for example, the second half of the chain; typically one would not want to use all the samples to exclude the burn-in region), and the new matrix is used as a subsequent input for the proposal density.

One of the most critical issues in MCMC sampling is the assessment of {\it convergence}, namely determining whether the probability distribution inferred from the chains resembles the true one. Two key requirements are that the entire relevant region of parameter space has been explored, and that the target distribution from the samples has become stable (\ie it would not change upon further sampling). The first point is relevant in particular in the presence of a multi-modal distribution; while in general the MCMC sampling should be able to correctly explore such spaces, it is wise to run multiple chains starting from different locations of parameter space, and we adopt this approach. Examples of algorithms that test convergence on single and multiple chains respectively are the Raftery and Lewis (hereafter RL) statistics (\citealt{Raftery&Lewis}) and the Gelman and Rubin ``R" test, comparing the variance of the mean within and between chains \citep{GelRu}.
All these tests are routinely performed by the publicly available software GetDist \citep{Lewis:2002ah}, which we use to analyze the chains. 
Convergence tests are meant to be performed on an unchanging proposal density. When the adaptive covariance matrix option is selected, the user can choose to stop changing the proposal density either when the desired acceptance rate is achieved (the default value is 25\%), or after a certain number of consecutive attempts to update the proposal density without a significant improvement of the acceptance rate. GalMC records the numbers of steps taken up to this time, so that they can be discarded in the computation of the posterior probability and in convergence tests.

For the analysis of data on the LAEs presented in this work, we use the information from the RL statistics to discard the burn-in of the chain and to obtain independent samples by thinning the chain by an appropriate factor. To ensure that convergence has been reached, we require the rather stringent constraint from the R test R$ -1 <$ 0.02 and follow the other general guidelines given in \cite{2005MNRAS.356..925D}.

\subsection{From parameters to observables}
\label{sub:like}

To compute the likelihood associated to a point in parameter space (in Bayesian terms, the likelihood of the data, given a model), we need to predict how the observed data would change as a function of the galaxies' physical properties that we want to measure. The parameter space we have so far explored is defined by the age since the onset of star formation, Age, the total stellar mass processed into stars, $M_{\rm tot}$, the dust reddening, defined by the excess color E(B-V), the metallicity in units of the Solar one, $Z/Z_\odot$, the redshift, $z$, and the e-folding time $\tau$ for exponential star formation history models. 
 In the analysis presented in this paper the redshift has been fixed, since it is well determined for narrow-band selected LAEs. Any combination of SED parameters can be explored, with the exception of $\tau$ and metallicity. Varying each of these parameters requires the use of a library of templates, and combining them requires too high memory usage to be convenient for an ordinary laptop computer. This does not affect our analysis significantly since our data do not show a strong preference for ESF vs CSF. Memory usage optimization for this purpose will be explored in a subsequent paper.

\subsubsection{Stellar population synthesis models}
\label{sec:sps}
As a starting point, one can use stellar population synthesis (SPS) models that predict the rest-frame flux as a function of wavelength for different ages, masses, metallicities, and star formation histories. We use the latest version of the publicly available GALAXEV code (\citealt{BCprivatecomm}, private communication; hereafter CB10), although GalMC also supports the earlier version of the same code (\citealt{2003MNRAS.344.1000B}, hereafter BC03). The main difference between the two versions is due to the different treatment of the TP-AGB evolution of low- and intermediate-mass stars, whose contribution produces redder rest-frame near-IR colors in the BC10 models (S. Charlot 2010, private communication). We modified the source code of  CB10 to reduce running time, which is essential for MCMC algorithms since tens of thousands 
of iterations are typically required. The main changes include a significant reduction
of output written to files by the program ${\rm csp\_galaxev}$ and the introduction of a new analytic option for exponentially increasing star formation rate (previously available only through the tabulated star formation rate option, which is considerably slower).

\subsubsection{Star formation history}

Three classes of star formation histories (SFH) can be specified: simple stellar population (SSP), corresponding to an instantaneous burst; constant star formation rate (CSF), and exponential star formation rate (ESF) $\psi(t) = 1/|\tau| e^{- t/\tau}$. In the latter case, the e-folding time  is also a parameter of the SED fitting, and is allowed to be either positive or negative to represent exponentially decreasing or increasing star formation histories. For the galaxy samples considered in this paper, the allowed choices of star formation histories allow us to find good fits to the data, but in general it would be ideal to measure the star formation history of a galaxy, rather than assuming a functional form for it. 
This method is employed, \eg, by the algorithms MOPED \citep{2000MNRAS.317..965H} and VESPA \citep{2007MNRAS.381.1252T}, where the star formation rate is recovered from the data in several age bins, and the data are also used to constrain the number of underlying stellar populations. We plan to explore this possibility in future application of GalMC, since we expect that high S/N data are needed in order to obtain meaningful constraints on the SFH.

We minimized the number of calls to ${\rm csp\_galaxev}$, which is computationally expensive, by introducing an ``adaptive library'' for exponential star formation history models. 
The user can specify the location of a library where the model files are stored; at each step the value of $\tau$ is compared to those of the models available in the library, and csp\_galaxev is run only if no close enough model is present. For the analysis in this paper, we have required an accuracy in $\tau$ of min(2\%, 2 Myr), but the criterion can be selected by the user. Each time a new model is computed, it is automatically stored in the library. The final product of the call to csp\_galaxev is a file containing the model as a function of age and star formation history; if required, the code also outputs a file containing the evolution of the stellar mass of the galaxy and a file containing the number of Lyman photons as a function of time. These files are used as input by the program galaxev\_pl, which extracts the spectrum at the relevant age and normalizes it to the chosen mass value. Thanks to these modifications, the average time per iteration of the GALAXEV part of our MCMC code is $\sim$ 0.3 seconds on a 2.66 GHz MacBook Pro, a factor of $\sim$ 20 faster than the publicly available GALAXEV code. 

\subsubsection{Metallicity}
The library of models available through GALAXEV comprise seven metallicity values, between $Z/Z_\odot$ = 0.005 and $Z/Z_\odot$ = 5. We allow the user to select a fixed value of metallicity different from any of the templates, or to include metallicity as one of the SED fit parameters. To compute the model spectrum for any value of $Z$, we interpolate between the two bracketing values available in the CB10 or BC03 stellar libraries. We tested both linear and logarithmic interpolation, finding that logarithmic interpolation (which became our method of choice) is more reliable. Our MCMC SED fitting runs are technically sound; yet, some caveats need to be mentioned. The dependence of an SED on metallicity is by its own nature complicated, since different types of stars contribute in different ways and at different epochs. Therefore, the precision of our SED modeling is necessarily limited by the paucity of empirical templates, and this systematic uncertainty should be folded into any metallicity measurement coming from photometric data relying on the same templates. We tested the magnitude of this effect by interpolating between two template values to find the predicted spectrum corresponding to one of the other template values, and found discrepancies of order $10\% - 20\%$, significant yet not unreasonable. 

We also explored the effect of using different priors on the metallicity distribution. Our choice was of a uniform prior in $\log Z/Z_\odot$, motivated by the observed distribution for Damped Ly-$\alpha$ systems \citep{2003ApJ...595L...9P}; however we also performed the MCMC analysis using a uniform prior in $Z/Z_\odot$ and found a mild dependence of our results on the prior used for the sample with less signal-to-noise. This is unsurprising since in general the choice of priors has more influence when the data have less constraining power. We further discuss this issue in Sec. \ref{sec:syst}.

\subsubsection{Impact of nebular emission}
\label{sec:neb}
If desired, the contribution to the model fluxes from nebular emission (from both continuum and lines) can be included. The strength of both continuum and lines is assumed to be proportional to the rate of H-ionizing photons per second, $Q_0$. 
Following \cite{1998ApJ...497..618S}, we describe the flux from the continuum emission as
\be
f_\lambda = \frac{c}{\lambda^2} \frac{\gamma(\lambda)}{\alpha_B} e_\gamma Q_0
\ee
where c is the speed of light, $\alpha_B$ is the case B recombination coefficient for hydrogen, $1 - e_\gamma$ is the escape fraction of Lyman photons, and $\gamma$ is the total continuum emission coefficient. After choosing a nominal electron density $N_e$ and temperature $T_e$ of the emitting gas, and a nominal helium to hydrogen ratio, the emission coefficients of the free-free, free-bound and two-photons continuum for hydrogen and helium can be found in \cite{1984ASSL..112.....A} and \cite{1980PASP...92..596F}. Following \cite{2009A&A...502..423S}, we use the values $e_\gamma$ = 1, $N_e = 100 \rm{cm}^{-3}$, $T_e = 10000 K$, and [He/H] = -1 throughout the paper; similar values were also recently measured for a star-forming galaxy at $z$ $\sim$ 2 by \cite{2011AAS...21711404R}. The corresponding template is added to the reference spectrum. We have checked that the evolution of the number of Lyman photons and the total emission computed in this way are consistent with those output by the publicly available code for computation of ionizing fluxes StarBurst99 \citep{1999ApJS..123....3L,2005ApJ...621..695V}.
To add the contribution of emission lines, we assume that the luminosity of the H$\beta$ line is given by
\be
L(H\beta) = 4.76 \times 10^{-13} e_\gamma Q_0
\ee
and we use empirical relative line intensities for H, He, C, N, O, S as a function of metallicity (D. Schaerer, private communication and \citealt{2009A&A...502..423S}; data from \citealt{2003A&A...401.1063A} and \citealt{1995MNRAS.272...41S}). This simple treatment is unlikely to capture the complex radiative transfer physics of the Lyman-$\alpha$ line. Since the narrow-band technique allows one to compute the real Lyman-$\alpha$ flux (as the line excess with respect to the continuum), L08 subtracted its contribution from the data and therefore we do not include it in our nebular emission templates.

\subsubsection{Galactic and IGM absorption}

The spectra thereby obtained can be corrected for absorption due to dust in the galaxy and to neutral hydrogen in the intergalactic medium (IGM). The former requires the assumption of a dust law to connect the SED fit parameter E(B-V) to an optical depth curve. GalMC currently supports two options: the Calzetti law \citep{1994ApJ...429..582C}, where the parameter $R_v$, related to the size and geometry of dust grains, can be chosen by the user, and a Milky-Way type law \citep{Cardelli:1989fp}. The former is used in the analysis of the present paper, with a value $R_v$ = 4.05 \citep{Calzetti2000}. Starlight absorption by the IGM is taken into account using the prescription from \cite{Madau:1995}. 

\subsubsection{Comparison with data points}

To obtain the spectra in the observed frame we compute the luminosity distance for the cosmology specified by the user through the present day matter density relative to the critical ($\Omega_{m,0}$) and the Hubble constant (H$_0$); a flat geometry ($\Omega_{\rm tot}$ = 1) is assumed. We use the luminosity distance to convert model luminosity to flux; the model fluxes are then convolved with the filter transmission curves specified by the user for each photometric band.  Finally, the flux densities $f_\nu$ 
are obtained as the number of photons corresponding to the fluxes of the model divided by the reference number of photons obtained for a flat spectrum of $f_\nu = 1 \,\mu$Jy:
\be
f^i_\nu = \frac{\int_{\lambda_{\rm min}}^{\lambda_{\rm max}} T^i(\lambda) f_\lambda(\lambda) \frac{c}{\lambda} d\lambda}{\int_{\lambda_{\rm min}}^{\lambda_{\rm max}} T(\lambda) \frac{c}{\lambda} d\lambda} 
\ee
where $T^i(\lambda)$ is the transmission curve in the $i-$th band between $\lambda_{\rm min}$ and $\lambda_{\rm max}$. The likelihood $\cal{L}$ for each model can now be computed (up to a normalization factor, which is irrelevant because we are interested in ratios of likelihoods) as a function of the $\chi^2$:
\be
- \log {\cal L} \propto \chi^2 = \sum_i \frac{(f^i_\nu - \phi^i)^2}{\sigma_i^2}
\ee
where $\phi^i$ is the flux in the $i$-th data point and $\sigma_i$ is its photometric error. 

\section{Algorithm testing}
\label{sec:test}
In this section we present the extensive tests we performed to ensure that our MCMC SED fit procedure produces reliable results. This goal can be achieved by generating mock galaxy catalogs with known physical properties and checking that the input parameters are correctly recovered by means of SED fitting. It is necessary to show not only that the estimates of the expectation values of parameters are unbiased, but also that the associated uncertainties, often quoted in terms of 68\% and 95\% confidence levels, are correct. 

Mock catalogs are built in the following manner. We use the modified BC10 SPS code to generate the spectrum of a mock galaxy in 13 bands from observed-frame UV to observed-frame IR and follow the procedure described in Sec. \ref{sub:like} to obtain the corresponding flux density in each band. To mimic a photometric uncertainty of 5\% in each band, we add a random Gaussian noise of this $1\sigma$ amplitude to the fluxes. While these assumptions on the uncertainty in the flux and the extent of the available photometry are realistic for our stacked fluxes of LAEs  \citep{2007ApJ...671..278G,2008ApJ...674...70L,Guaita2010}, the results of this test do not depend on the particular numbers. 

We build and test mock catalogs for both constant and exponential star formation histories. For the constant star formation case, the running time of the MCMC SED fitting code is limited; a chain of composed by a few thousands steps can be obtained in approximately two hours on a 2.66 GHz MacBook Pro laptop computer, and we have found that in this simple case, when three parameters describe the SED, this is enough to achieve convergence. This makes the CSF scenario a suitable test case to check our marginalized probability distribution of parameters, since this task requires running GalMC on a large number of Mock catalogs, as explained below. Mock catalogs with exponential star formation history are used as a means to test robustness to the presence of degeneracies in the SED and the effect of using different variables and priors in describing the star formation history. Depending on the sign and ratio of age and $\tau$, the timescale associated to the exponential rise or decline of the SFR in the mock model, the posterior probability of parameters can be multi-modal or even flat if true degeneracies are present. 
By testing the MCMC SED fit in such scenarios, we ensured that the input parameters are correctly recovered even in the presence of degeneracies and that we don't overestimate the degree of belief in our results (for example, by assuming that chains have converged when this is not the case). Furthermore, we explored different parameterizations in $\tau$ and were able to choose tFphe one that leads to the most reliable results. 

\subsection{Constant Star Formation Models: \\ Parameter and error estimation recovery}
\label{sec:mockCSF}
Our reference model is a galaxy characterized by constant star formation history, Age = 180 Myr, total mass converted into stars M$_{\rm tot}$ = $2.95 \times 10^8$ M$_{\odot}$, and excess color E(B-V) = 0.2. We build 200 mock realizations of this galaxy, adding a Gaussian random scatter to each data point of $1\sigma$ amplitude equal to the 5\% photometric error. We run the MCMC SED fitting code on the reference model (without scatter in the photometry), ensuring that we correctly recover the input parameters, and hereby obtaining a reliable covariance matrix which gives a nearly optimal acceptance rate (around 30\%). We then run the MCMC SED fitting code on each of the mock catalogs, using this covariance matrix as an input for the proposal density. After discarding the chains that have not yet converged after 10000 steps, we are left with 172 reliable runs that we use to test the recovery of uncertainties. To do so, we compute the frequency with which each of the ``true" (input) parameters is within the 68\% and 95\% confidence levels from the 1-D marginalized posterior probability distributions obtained for the mock catalogs. The $i$-dimensional {\it marginalized} posterior distribution of parameters is obtained by integrating the $n$-dimensional posterior distribution $p$ in $n$-$i$ dimensions; for example, for CSF the 1-D age marginalized posterior probability is given by the function:
\be
p{\rm(Age}) = \int d{\rm M}_{\rm tot} \int d{\rm E} \; p {\rm(Age, M}_{\rm tot}, \rm{E)}. 
\ee
Since a thinned MCMC chain has the property that the density of points is proportional to the posterior probability (the product of likelihood and priors), the above integral is easily computed as a sum over the points of the chain that takes into account the time spent at each location in parameter space.
By definition, we expect that the true values are within the region allowed at 68 (95)\% confidence 68 (95)\% of the time. We find that this is the case for each parameter, within the Poisson fluctuation error (order of 8\% effect) associated with our statistics. 
We conclude that the uncertainties we report are reliable. We note that since we use the well-tested GetDist software to compute the posterior probability distributions and to perform convergence tests, this check ensures that our sampling algorithm and calculation of likelihood are implemented correctly. 

A more immediate visualization of the match between the observational errors and the corresponding uncertainties in the parameters can be obtained by plotting the probability distribution of the parameters in the ``true" model (without the scatter), and the scatter of the best-fit values obtained for the 172 mock catalogs. The agreement between these two quantities is however only expected in the case of perfect (one-to-one) correspondence between the likelihood associated to the data for each model and the posterior probability, and assuming that the best fit represents the truth for each model.
These assumptions are not unreasonable for these simple models, leading to the observed agreement shown in Fig. \ref{fig:errors}.

\subsection{Exponential Star Formation Models: Parameter recovery}
\label{sec:mockEFSH}

\begin{figure}[t]
\begin{center}
\includegraphics{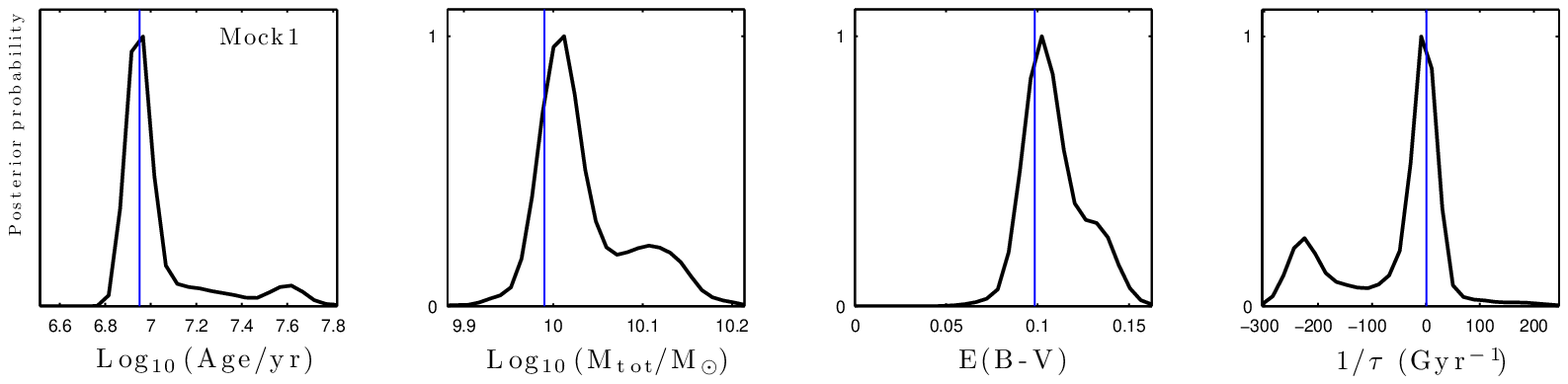}
\includegraphics{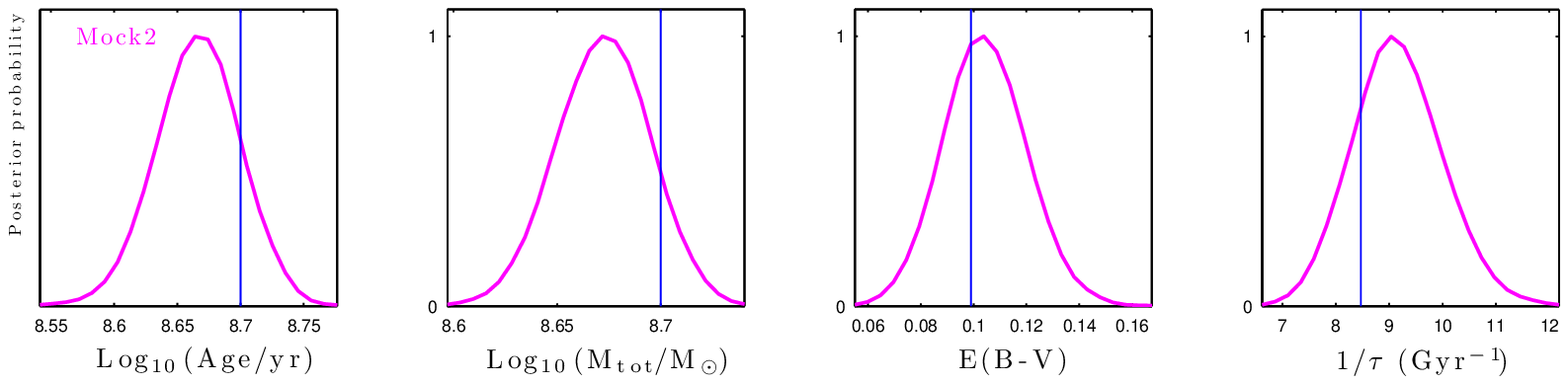}
\includegraphics{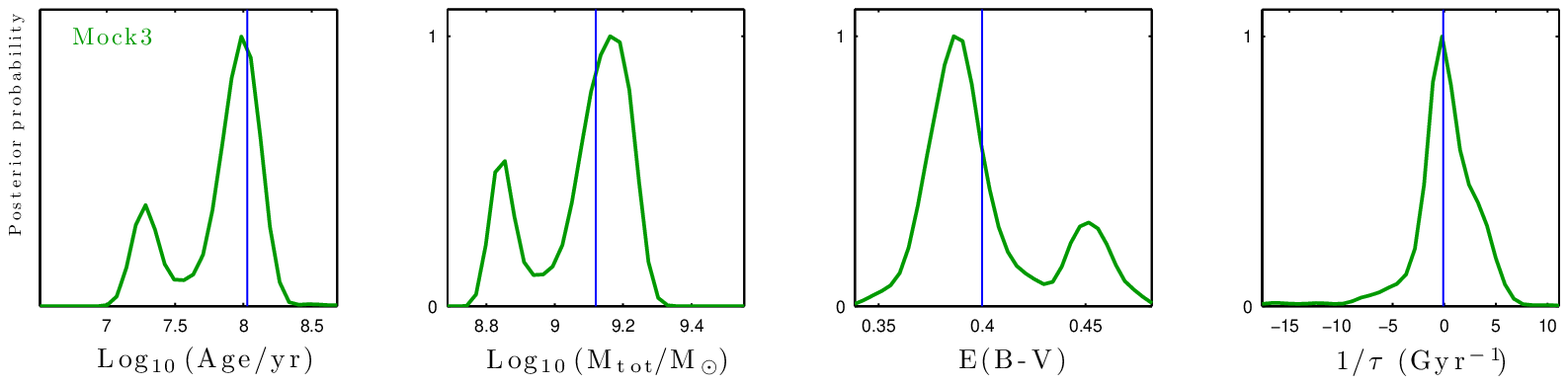}
\caption{Each row shows the marginalized 1-D posterior probability distribution for the SED fitting parameters for one of three illustrative mock catalogs of galaxies with Exponential Star Formation Rate (increasing and decreasing). As discussed in the text, multi-modality in the posterior is often observed for such models. The input parameters are shown as blue vertical lines; in each case the agreement between input and recovered parameters is good.}
\label{fig:mockESFH}
\end{center}
\end{figure}

Models with exponentially decreasing star formation rate have been often used in the literature, and recently increasing exponential star formation rates have also been considered \citep{2010MNRAS.407..830M, Guaitaetal2010b, 2010ApJ...725.1644L}. 
While studying the constraints on these models is interesting, they are often affected by degeneracies intrinsic in the model SEDs. 
For this reason, it is necessary to pay particular attention to the use of suitable sampling variables and priors. We have implemented three possible choices of parametrization in $\tau$, using as sampling variable $\tau$, $1/|\tau|\ln(|\tau|)$, and $1/\tau$; all of these are currently available options in the code. We ran GalMC for each of these choices on a range of Mock catalogs, using different signs and numerical values for the ratio of age and $\tau$. The parametrization $1/\tau$ was our final variable of choice, since we did not observe any bias or erroneous convergence problem for this case. 
This description has the attractive feature that the constant star formation rate case, which is the limit for $\tau \rightarrow \infty$ for both positive and negative values of $\tau$, occupies a single spot ($1/\tau =0$) in parameter space (unlike the other parameterizations). This makes the interpretation of constraints easier and transparent. 
We implemented it with a flat prior in $\ln \tau$. In fact, during the parametrization selection process, we found that the use of a flat prior in $1/\tau$ led in some cases to a poor recovery of the input parameters,
as did the use of a flat prior in $\tau$ when sampling using $\tau$ as one of the parameters, which is not uncommon in the literature. On the other hand, no bias or erroneous results were found by us when using a uniform prior in $\ln |\tau|$, as shown in Fig. \ref{fig:mockESFH}, confirming that this is the best choice of prior.

The sensitivity of the results to the choice of priors emphasizes the need for caution when computing constraints on these models. We expect that significant improvement could be achieved by using variables which are closer to the eigenvectors of the covariance matrix; we defer this treatment to a subsequent paper (Acquaviva et al 2011 in prep).

Results obtained for three illustrative mock catalogs, chosen to have different star formation histories (increasing and decreasing), $\tau$/Age ratios, and redshifts, are shown in Fig. \ref{fig:mockESFH}. The 1-D posterior distributions are obtained by integrating the posterior probability over the remaining N-1 dimensions. In all the Figures, the normalization of these plots is arbitrary; we show it using the usual CosmoMC convention, where the height of the peak is one. 
We find good agreement, in each case within the region allowed at 68\% confidence, between the input parameters and those recovered by means of SED fitting. \footnote{For Mock catalogs 1 and 3 convergence is slow and the stringent criterion $R-1 < 0.02$ would not be satisfied yet. This is encouraging since it indicates that our criteria are indeed fairly conservative.}

\subsection{Comparison with grid-based techniques}
\begin{figure}[t]
\begin{center}
\includegraphics[width=\linewidth]{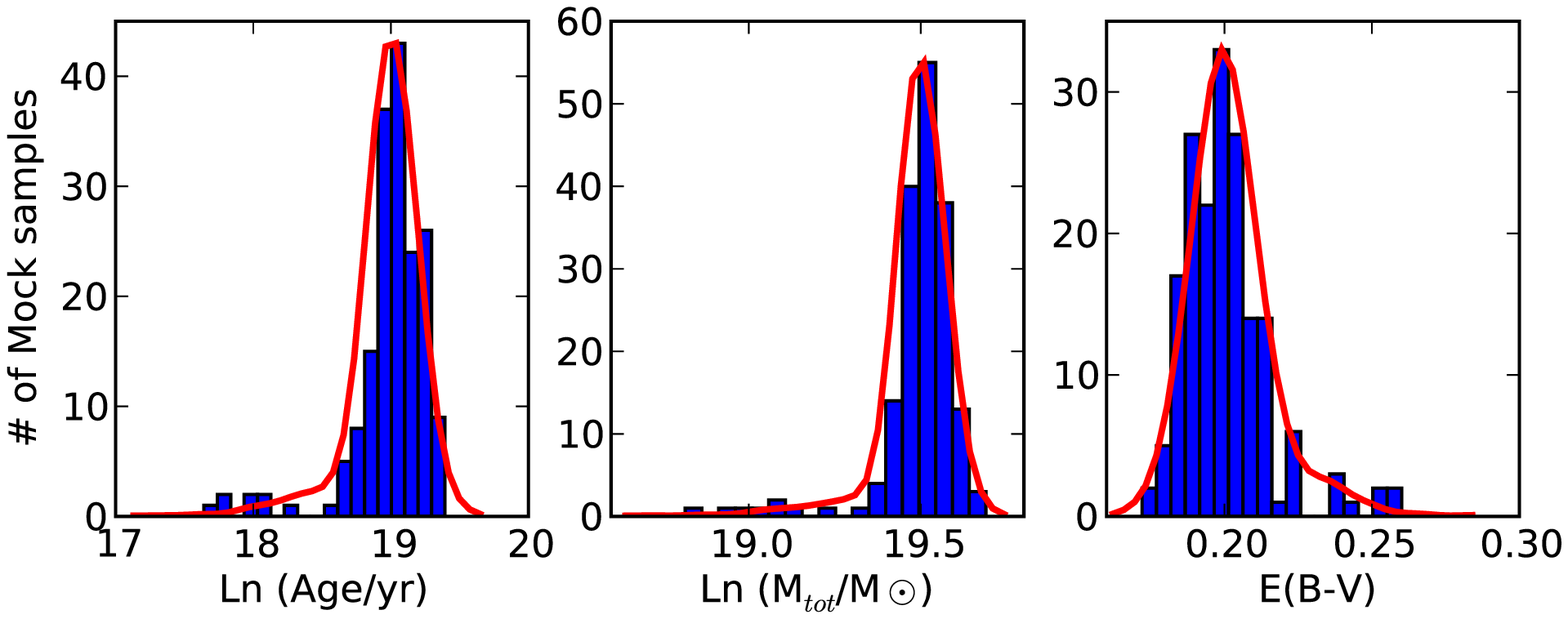}
\caption{The red line shows the posterior probability distribution for the reference model described in the text. The blue histogram represents the distribution of best-fit values obtained from the analysis of 172 Mock data realizations, obtained convolving the reference fluxes with a random Gaussian scatter of amplitude equal to the photometric error of the reference model. The agreement confirms the reliability of our error estimation procedure.}
\vspace*{10cm}
\label{fig:errors}
\end{center}
\end{figure}

As explained in Sec. \ref{sec:MCMC}, the MCMC technique allows one to draw samples from the posterior distribution, and to use them to compute any meaningful quantity associated with it, \eg the expectation values of parameters, and their uncertainties (for any quantile). This is obviously not the only way to achieve this goal.
 Why should one make use of the MCMC technique rather than sample the posterior distribution by means, for example, of a fine grid in the parameter space? There are two main arguments in favor of this choice: {\it efficiency} and {\it reliability}. The first can be understood as follows.
In general, the allowed range of values for each parameter is much larger than the interesting range for that parameter, namely the region where the likelihood is significantly different from zero, which is the relevant one in computing integrals of the form given by Eq. (\ref{eq:prob}). For example, to measure the age of a galaxy, one would typically want to test values from the youngest age that be can resolved (possibly a few million years) to the age of the Universe at the redshift of the galaxy. Even using a logarithmic spacing, this spans several e-foldings. On the other hand, the width of the interesting region (say, to within 99\% confidence level from the best-fit or mean age) will typically be much smaller (of the order of three e-foldings for the example of Fig. \ref{fig:errors}). By sampling on a grid, a large fraction of time will be spent sampling regions of no interest, while in MCMC, after a burn-in period, every step, whether or not is accepted, is taken in the informative region. For an equal amount of CPU time, this means that the finesse of the results achieved by MCMC - the accuracy when computing \eg the integral in Eq. (\ref{eq:prob}) - is much higher. As an example, we considered the Mock catalog described in \ref{sec:mockCSF}. 
We found that in $86\%$ of our 200 test runs, 10000 steps were enough to achieve convergence. Let us assume, to be conservative, that 20000 steps is a safe number of steps for a hypothetical MCMC chain to converge. For this three-dimensional parameter space, this total number corresponds to computing the model SED for $\sim$ 27 values for each parameter. We performed the grid sampling assuming an allowed range which is comparable for $\log$(Age) and E(B-V) to the top-hat priors used for the MCMC case, and significantly narrower in $\log$(Mass). The number of points falling in the relevant region of the posterior (shown in Fig. \ref{fig:errors}) is only a handful (three or four) for $\log$(Mass) and E(B-V), and about ten for $\log$(Age). Therefore, the fraction of time spent in the informative region is only $(4 \times 4 \times10)$/$27^3$, which is less than $1\%$. This is too coarse a grid to reconstruct the marginalized probability distribution, and an attempt at it via the integration of the posterior leads to a delta function at the best fit found in the grid ($\chi^2 = 20$, Age =108 Myr, Mass = $2.62 \times 10^8$ M$_{\odot}$, and E(B-V) = 0.23, to be compared with the input values Age = 180 Myr, Mass = $2.95 \times 10^8$ M$_{\odot}$, and E(B-V) = 0.2).
 The situation only worsens for spaces with higher dimensionality, since the number of MCMC steps typically scales linearly in the number of parameters $N$, while the number of steps in grid-based methods grows exponentially. For six parameters, the expected numbers of MCMC steps would be of the order of $6 \times 10^4$, which corresponds to sampling less than 10 values for each parameter on a grid in the same CPU time.
 
Reliability is an equally important issue. The spacing on a grid has to be chosen arbitrarily (and, as we saw, it is computationally very expensive to choose a fine pace), and there is no guarantee that one would not miss features of the posterior distribution which are narrower than the interval between adjacent values. Conversely, in MCMC one can use an adaptive step size (such as the one implemented by us by means of the adaptive covariance matrix), sampling more finely in high-likelihood regions. Furthermore, the use of convergence tests and the possibility of running multiple chains provide strong indications that all the interesting regions of parameter space have been adequately explored. These attractive features of MCMC technique allow for a true optimization of CPU time.

\section{Results}
\label{sec:results}
\subsection{LAE samples}
As part of the MUSYC survey \citep{2006ApJS..162....1G}, \cite{2007ApJ...667...79G} discovered a complete sample of 162 Lyman Alpha Emitters (LAE) at $z$=3.1 in a narrowband survey of the Extended Chandra Deep Field South (ECDF-S). The available observed-frame broad-band photometry for this sample encompasses six UV-optical bands (U, B, V, R, I, z), two near-IR bands (J and K), and the four IR IRAC bands. 
L08 eliminated from this sample 86 LAEs in regions of the IRAC images suffering significant contamination from nearby neighbors, and created samples of 18 IRAC-Detected and 52 IRAC-Undetected LAEs, with the ``detection'' at flux density
$\geq 0.3 \mu$ Jy equating to roughly 3$\sigma$ significance. Six additional IRAC detections were classified as probable AGNs or high-dust galaxies and were analyzed separately.
Median-stacked SEDs of these samples were formed and fit using BC03 models; for constant star formation rate, 
best-fit models showed no dust for
either sample and stellar masses of $9\times10^9$ [$3\times10^8$] M$_\odot$ and ages of 1600 [160] Myr for the IRAC-Detected [Undetected] sample; the corresponding uncertainties at 68\% confidence are reported in Table \ref{tab:syst}. The best-fit values obtained for exponentially declining SFH were within 20\% of the values obtained for CSF.
\cite{2007ApJ...671..278G} fit a two-population SED model to the stacked IRAC-Undetected SED, using an exponentially declining SFH for each population.
Although their best-fit model placed 80\% of the stellar mass in an underlying old population and only 20\% in a $\sim$ 20 Myr-old starburst population, they were unable to rule out a single-population fit with age $\sim$ 150 Myr, in good agreement with the results of L08.
In the following, we investigate further the median-stacked SEDs of the L08 IRAC-Detected and IRAC-Undetected samples of $z=3.1$ LAEs;
we refer to these two samples as ``IRAC Det" and ``IRAC Und" in figures and tables.
Our treatment differs from the previous analyses in the use of the full probability distribution of parameters obtained from the MCMC SED fitting code, as well as in the inclusion of the contribution of nebular continuum and emission lines and in the use of the improved CB10 SPS models.

\subsection{Physical properties and SFH}
We begin our MCMC analysis considering a four-dimensional parameter space defined by age, dust reddening, metallicity, and stellar mass, and assuming constant star formation history. The dust reddening is parametrized by the excess color E(B-V) assuming a Calzetti dust absorption law. The input parameter of our modified GALAXEV code is the total mass of gas turned into stars M$_{\rm tot}$ (the integral of the instantaneous star formation rate over the age of the galaxy), but we report constraints on the more meaningful stellar mass M$_*$, which takes into account the life cycle of stars and the associated mass loss. This effect is typically of order of $10-20\%$. We assume that none of the gas thereby injected into the IGM is reprocessed to stars. We use top-hat priors on $\log$(Age), $\log$(M$_{\rm tot})$, $\log_{10} (Z/Z_\odot)$ and E(B-V); the allowed ranges are reported in Table \ref{tab:param}. The lower bound of $10^6$ years for the age comes from an educated guess of the applicability of the SPS models we consider \citep{Bruzual:2003}.
Our reference cosmology assumes total energy density relative to critical $\Omega_{tot} = 1$, matter density $\Omega_m = 0.258$, and Hubble constant $H_0 = 73$ km/sec/Mpc.
In our baseline model we assume a Salpeter Initial Mass Function (IMF) as in L08 
and we include the contribution of nebular emission as described in Sec. \ref{sec:neb}. While in simple stellar population  (SSP) models the contribution of emission lines and continuum become negligible after $\sim$ 20 and $\sim$ 5 Myr respectively (\eg \citealt{2010arXiv1009.0007N}), when star formation is ongoing Ly-$\alpha$ photons are continuously produced and this effect cannot be neglected. For example, in CSF models the number of Ly-$\alpha$ photons increases for the the first $\simeq$ 20 Myr and stays roughly constant thereafter. It is however true that for older stellar populations the relative importance of nebular emission with respect to the stellar continuum decreases with time. 

\begin{figure}[p]
\begin{center}
\includegraphics[width=\linewidth]{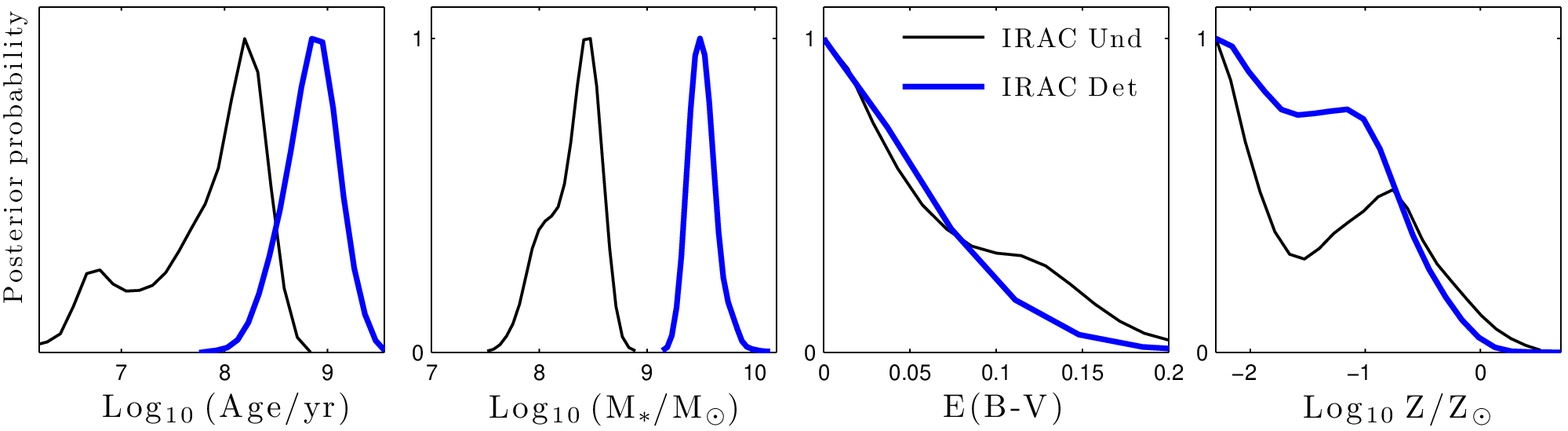}
\includegraphics[width=\linewidth]{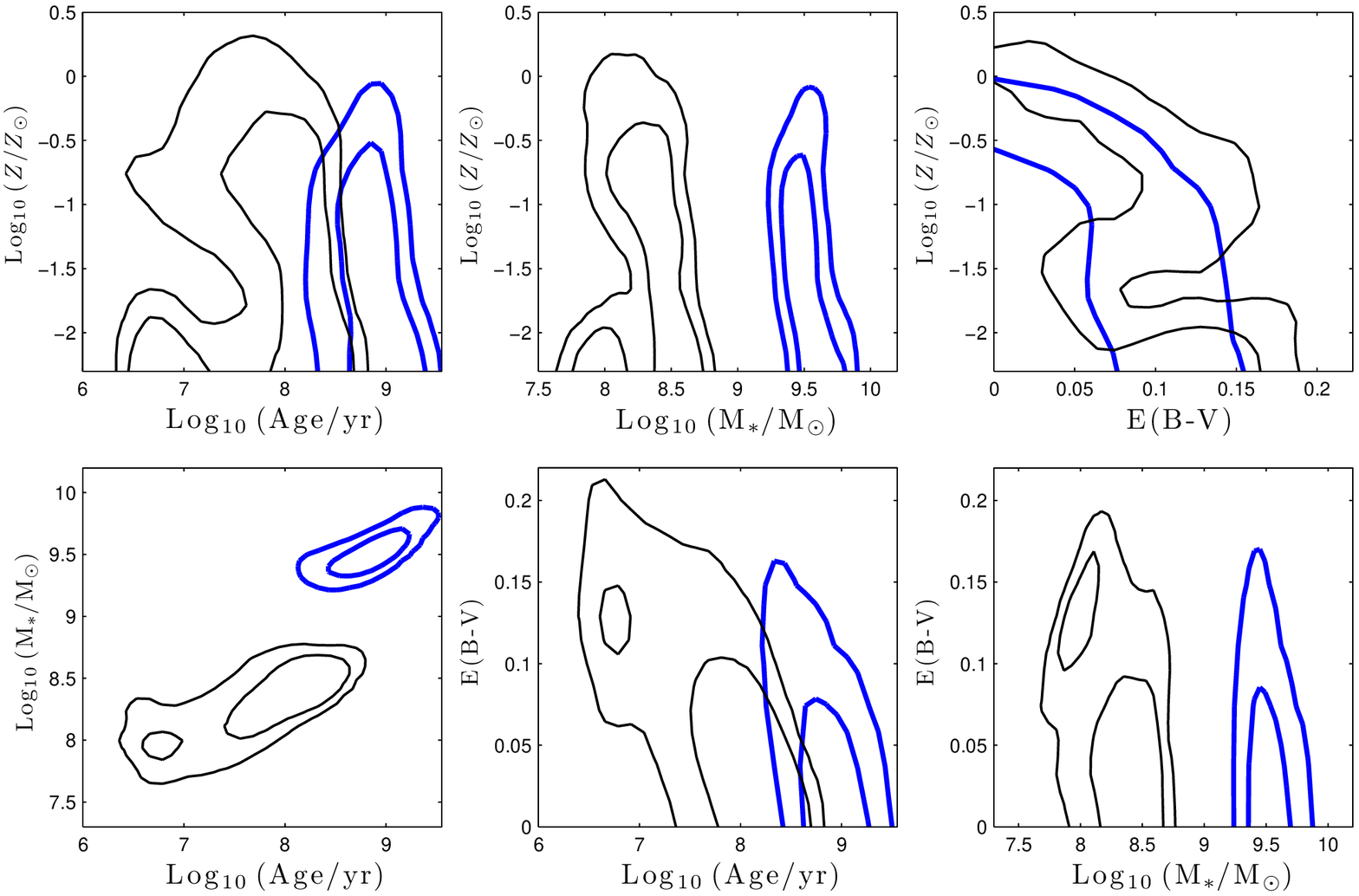}
\caption{{\bf Top row}: 1D marginalized posterior distributions for constant star formation history models, for the IRAC-undetected (black, thin) and IRAC-detected (blue, thick) stacked samples. {\bf Bottom rows}: 2D marginalized contours defining the 68 and 95\% confidence regions on the corresponding pairs of parameters.}
\label{fig:Constraints_CSF}
\end{center}
\end{figure}

Results for the IRAC Detected and Undetected samples are presented in Fig. \ref{fig:Constraints_CSF} and summarized in Table \ref{tab:results}. 
 In the table we report the mean value of parameters from the posterior distribution $p$, \eg:
\be 
< {\rm Age} > = \int d {\rm Age} \int d{\rm M}_{\rm tot} \int d {\rm E  \; Age} \; p{\rm (Age, M}_{\rm tot}, {\rm E}) 
\ee
where the integral is over the domain specified above. This is easily computed by averaging over the R samples $r_i$ obtained after thinning the Markov chain:
\bea 
< {\rm Age} > & = & \int d {\rm Age} \int d{\rm M}_{\rm tot} \int d {\rm E  \; Age} \; p{\rm (Age, M}_{\rm tot}, {\rm E}) \nonumber \\
&& \simeq \frac{1}{R}\, \sum_i \,{\rm Age(}r_i)
\eea

 If the posterior is not symmetric (as in this case), the mean values can differ from the best-fit values; we return to this issue in Sec. \ref{sec:syst}. We also report the $68\%$ uncertainties, computed integrating the posterior from each side toward the high-probability region and stopping when the integral under the curve on each side is $16\%$ of the total (32\% in the case of E(B-V), which has a one-tail distribution). In Fig. \ref{fig:Constraints_CSF} we show 1-D the marginalized posterior probability distribution for each parameter, and the 2-D marginalized constraints, including contours for the $68$ and 95$\%$ confidence regions, on the different combinations. Degeneracies between parameters appear here as diagonal axes of these ellipse-like curves; the one between age and stellar mass is evident in the bottom left panel. 

The different nature of the two samples is easily seen from Fig. \ref{fig:Constraints_CSF}. While both stellar populations are consistent with having no or very little dust  (E(B-V) $<$ 0.04), the sample detected in IRAC is considerably older and more massive than the undetected one, with mean ages and stellar masses of 0.67 [0.37 - 1.18] Gyrs and $3.2 [2.5 - 4.2] \times10^9$ M$_\odot$ versus 6 [1-20] $\times 10^7$ yrs and 2 [1.1 - 3.4] $\times 10^8$ M$_\odot$ respectively. We are also able to set a constraint on the metallicity of these LAEs. For the IRAC detected sample, we find $Z/Z_\odot$ = 0.036 [0.005-0.07], while for the IRAC undetected sample,  $Z/Z_\odot$ = 0.05 [0.005-0.13]. In both cases, Solar metallicity is excluded at more than 95\% confidence. We note however that using a different prior (uniform in $Z/Z_\odot$) slightly relaxes the allowed parameter range. The metallicity constraint becomes weaker for the IRAC undetected sample, which has lower S/N; in this case Solar metallicity is only excluded at 68\% confidence. This interesting result is in alignment with previous claims
that $z$=3.1 LAEs have metallicity lower than Solar \citep{2011ApJ...729..140F} and might be galaxies in their first star formation episode (\eg \citealt{1998ApJ...502L..99H}).

\begin{figure}[p]
\begin{center}
\includegraphics[width=\linewidth]{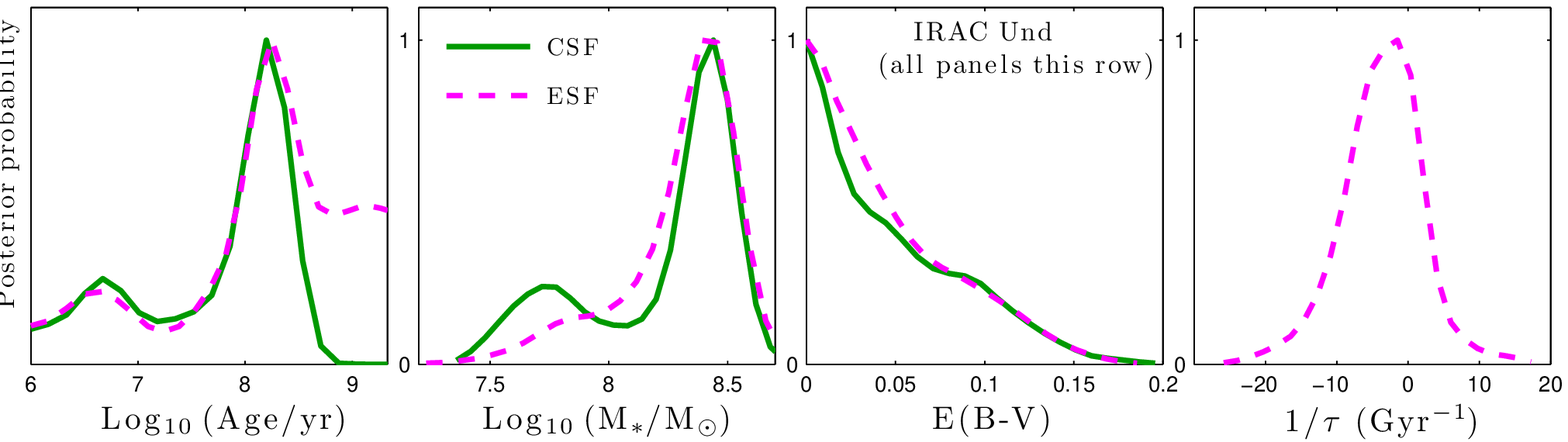}
\vspace{0.5cm}
\includegraphics[width=\linewidth]{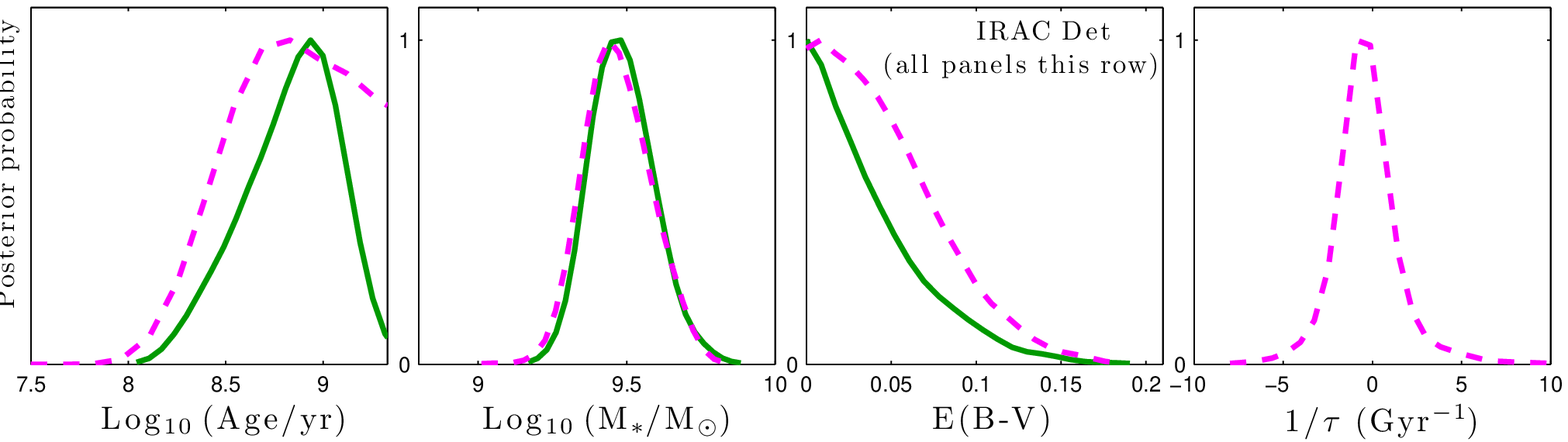}
\caption{{\bf Top row}: Impact of the assumption on the star formation history (constant, green, solid; exponential, magenta, dashed) on the posterior probability distribution of SED fit parameters. For this analysis we held the metallicity fixed at the value $Z = 0.02 \, Z_\odot$. The main appreciable difference is the inability to rule out very old ages (comparable to the age of the Universe at z = 3.1) for exponential SFH, for both the IRAC Undetected ({\it top}) and IRAC Detected ({\it bottom}) samples. Our constraints on $1/\tau$ show that for both samples the CSF value $1/\tau$ = 0 is well within the region allowed at $68\%$ confidence.}
\label{fig:Constraints_ESFH}
\end{center}
\end{figure}

We also consider the effect of assuming an exponential star formation history. In this case there is an additional parameter of the SED fit, the e-folding time $\tau$. As explained in Sec. \ref{sec:mockEFSH}, we use the variable $1/\tau$ in the MCMC sampling; we apply a flat prior in $\log|\tau|$ between the values of $\tau$ = -4 Gyr and $\tau$ = 4 Gyr (for $|\tau| > 2$ Gyr the SED is indistinguishable from that of a Constant Star Formation History, so that the CSF case is included by this parametrization as the limit of $1/\tau \rightarrow 0.25$ Gyr$^{-1}$). 
For this analysis, we fix the metallicity at the same value for both samples, in order to isolate the effect of assuming a more general SFH. We chose the value $Z/Z_\odot$ = 0.02, which is consistent with the range found earlier and for which we have an empirically calibrated stellar template available.
The results are shown in Fig. \ref{fig:Constraints_ESFH}; we consider the four most meaningful 2D combinations of parameters. We plot the posterior probability as function of $1/\tau$ to show the output of the MCMC SED fitting algorithm, but we report the constraints in terms of $\tau$ for clarity in Table \ref{tab:results}. 

In both cases the preferred value of $\tau$ is close to the CSF value. For the Undetected sample, we find $|\tau| >$ 0.12 Gyr at 68\% confidence, while for the Detected sample the corresponding constraint is $|\tau| >$ 0.67 Gyr. This can perhaps be interpreted in terms of the different age of these two stellar populations. In fact, the SED is sensitive to the ratio Age/$\tau$; when this ratio is small, there aren't enough e-foldings to distinguish the effect of the exponential SFH from a Constant one. 
The change in the probability distribution of the other parameters is also shown in Fig. \ref{fig:Constraints_ESFH}. The main effect of the inclusion of an extra parameter is the inability to rule out very old ages (leftmost panels), so that the sharp difference in the age of the two populations found for CSF is somewhat mitigated. The results for the stellar mass and dust reddening are stable to the change in SFH. This was observed previously by L08 for these $z$=3.1 samples and was also found to be true for redshift $z$=2.1 LAEs by \cite{Guaitaetal2010b}. 
Overall, both the IRAC Detected and Undetected LAEs are reasonably fit by a Constant Star Formation History model characterized by Age, stellar mass, and dust reddening E(B-V), with fixed metallicity. The best fit models for the two cases have a $\chi^2$ of 13.4 and and 7.6 respectively, for twelve data points and three free parameters. 
In the case of exponential SFH, the preferred $\tau$ values are very close to the ones corresponding to CSF, but there is a modest improvement in the goodness-of-fit coming from adding one extra parameter, with best-fit $\chi^2$ of 10.8 and and 5.2 respectively. These values are obtained for an exponentially increasing SFH. However, CSF is well within the range of e-folding times allowed at 68\% confidence for both samples, and we conclude that the inferred physical properties of our LAEs are robust to different assumptions on the SFH and that the CSF model is a satisfying parametrization for both samples.

\subsection{Impact of SPS modeling}
\label{sec:syst}
Besides the choice of star formation history, many features of the stellar population synthesis modeling influence the results of SED fitting. Our preferred model includes nebular emission as described in Sec. \ref{sec:neb}, considers metallicity $Z$ as one of the SED fit parameters, and employs the latest version of the GALAXEV code and templates,  CB10. Here we change these assumptions one by one in a sequence and show their effect on the probability distribution of SED parameters and therefore on the inferred properties of LAEs. We first fix the metallicity at the Solar value, then neglect the effect of nebular emission, and then use the earlier version of the GALAXEV code and templates, BC03. The final combination of assumptions coincides with those used in the analysis of L08. 

\begin{figure}[p]
\begin{center}
\includegraphics[width=\linewidth]{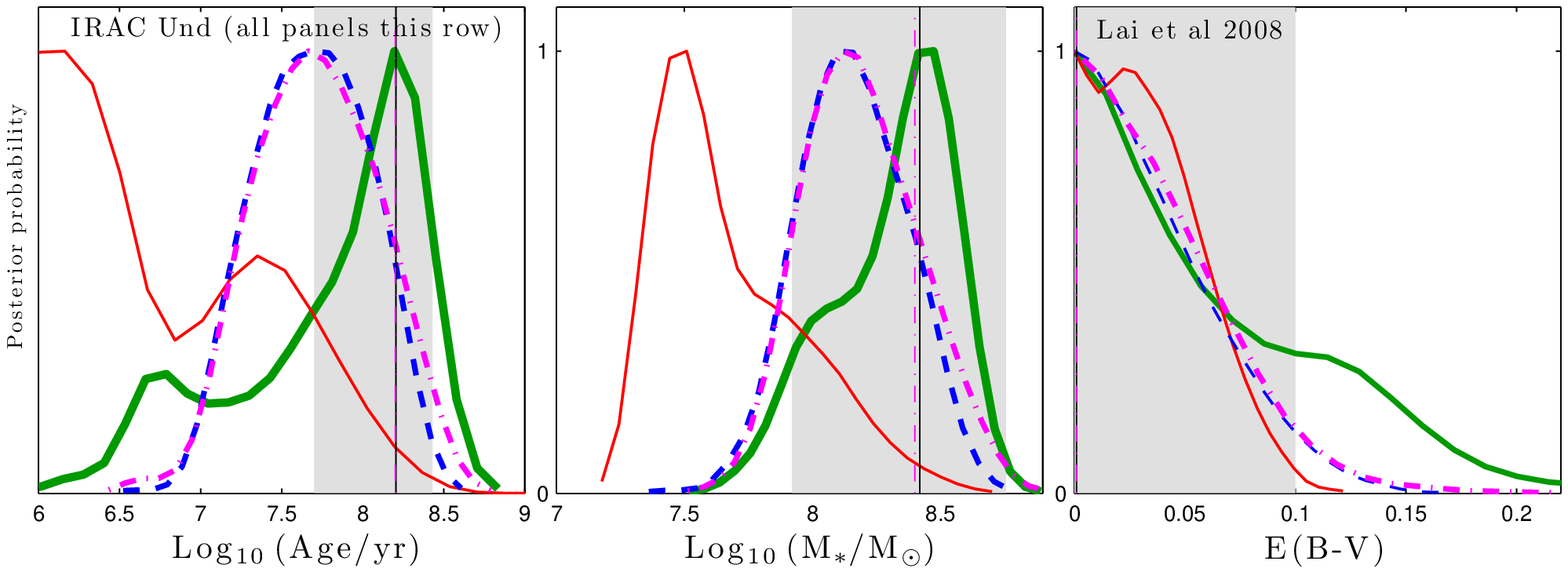}
\includegraphics[width=\linewidth]{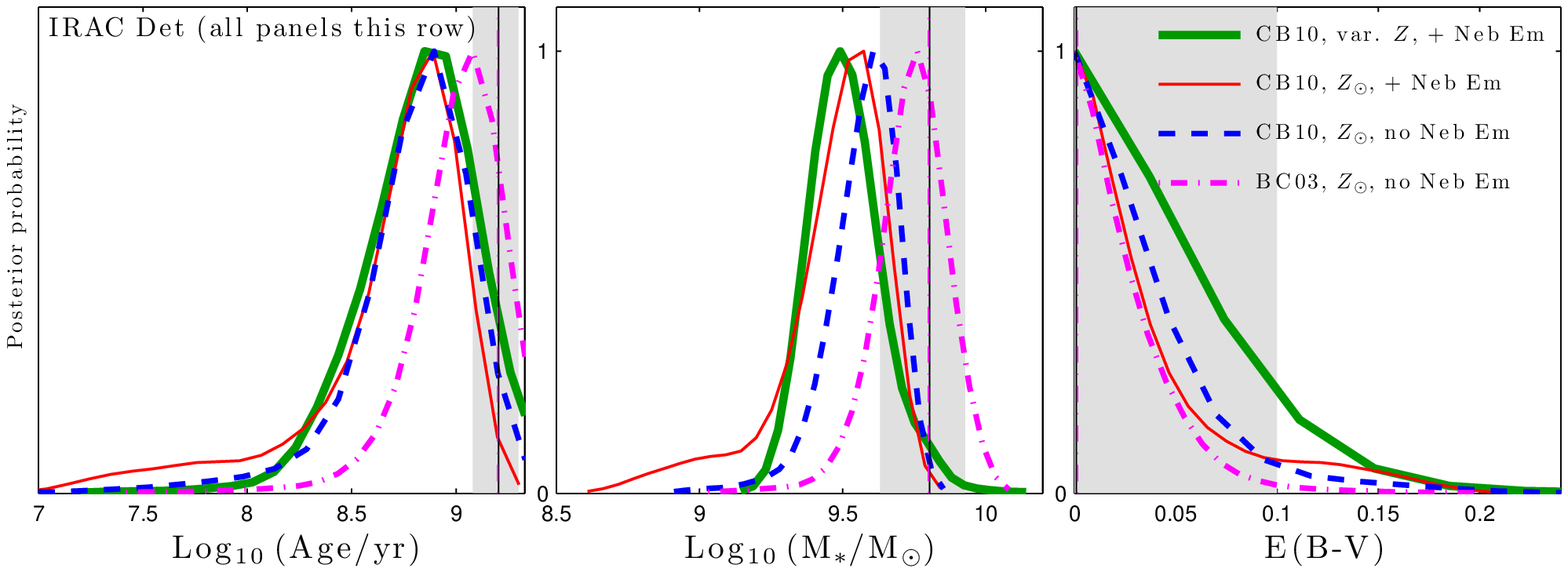}
\caption{Impact of inclusion of nebular emission, metallicity, evolution in SPS templates, and best-fit vs mean values on the posterior probability distribution of SED fit parameters for the IRAC Undetected ({\it top}) and IRAC Detected ({\it bottom}) samples. Constant Star Formation History is assumed. The thin black vertical line and the grey shaded area indicate the best-fit models and $68\%$ confidence levels from Lai et al 2008. The magenta dot-dashed vertical lines and curves show the best-fit models and probability distributions of our SED fit for the same SPS models and metallicity. This comparison shows the slight shift resulting from the use of mean values of the posterior distribution as opposed to best fit values. For the Undetected stack, the introduction of nebular emission causes the appearance of a second mode at very young ages and very low masses, which is particularly relevant at solar metallicity, where the contribution of the emission lines is stronger. Results are summarized in Table \ref{tab:syst}.   }
\label{fig:Systematics_All}

\end{center}
\end{figure}

Our results are shown in Fig. \ref{fig:Systematics_All} and summarized in Table \ref{tab:syst}; the last column also shows the best-fit $\chi^2$, for twelve data points and four (for varying $Z$) or three free parameters. In all cases the inferred amount of dust reddening is nearly insensitive to the assumptions on the SPS modeling, except for a moderate increase of the uncertainties when metallicity is varied, and we do not discuss it further. This behavior is expected because we subtract the Ly-$\alpha$ flux from our photometry before fitting. As a result, the impact of nebular emission lines and continuum, as well as the difference among the stellar templates we used, is negligible in the rest-UV region of the spectrum (observed frame UV-optical), whose slope measures dust reddening.

The effect of assuming fixed Solar metallicity, as in L08, produces an appreciable effect on the probability distribution of SED parameters for the IRAC Undetected sample. This same trend with metallicity was also observed by L08 for two discrete values of $Z/Z_\odot$; however, a direct comparison is not possible because we use the CB10 models where they used the BC03 ones. The improvement in the best-fit $\chi^2$ for models of low metallicity is especially marked for the IRAC Detected sample, as shown in Table \ref{tab:syst}. 

\begin{figure}[p]
\begin{center}
\includegraphics[width=8cm]{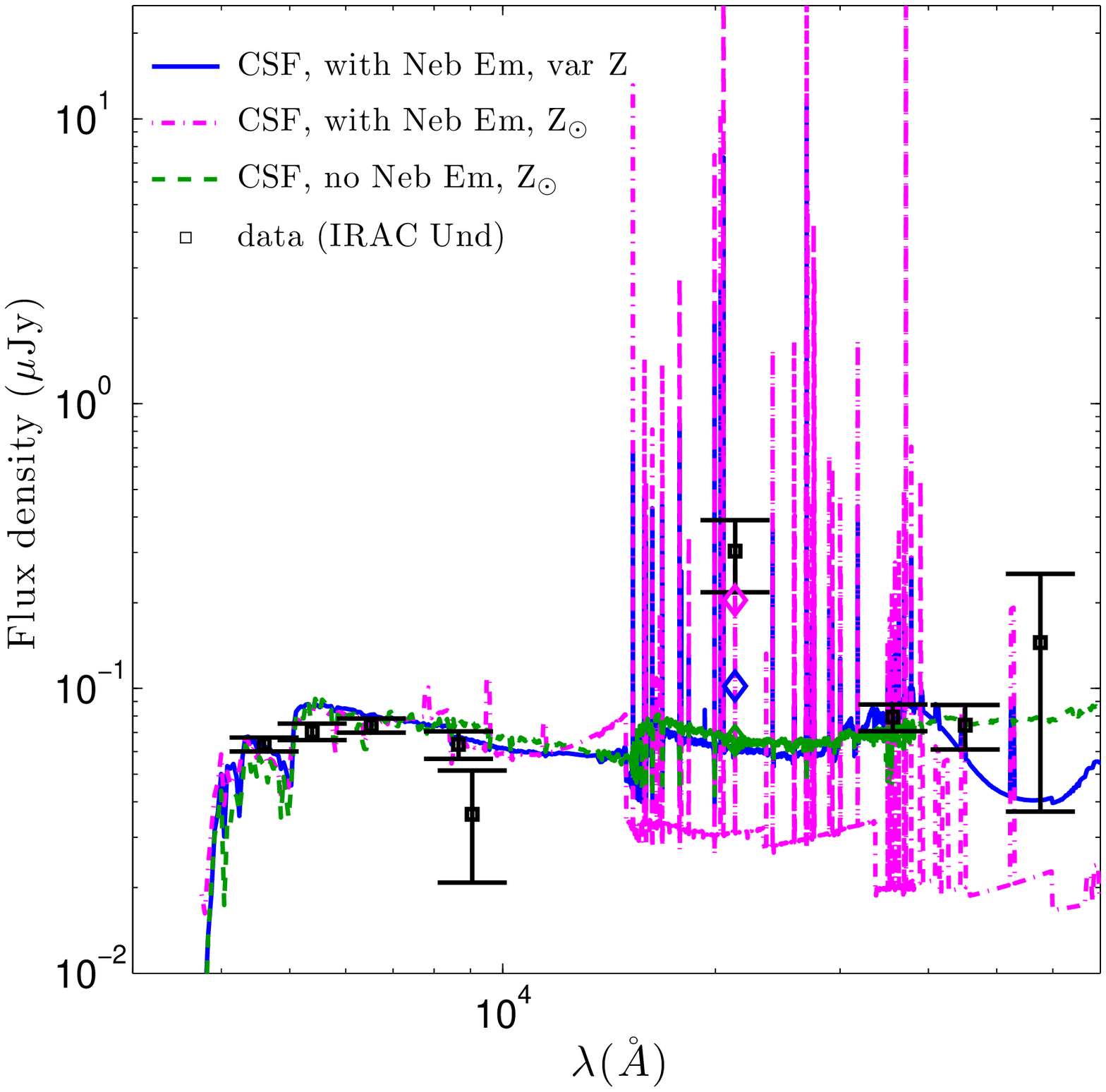}
\hspace{0.2cm}
\includegraphics[width=8cm]{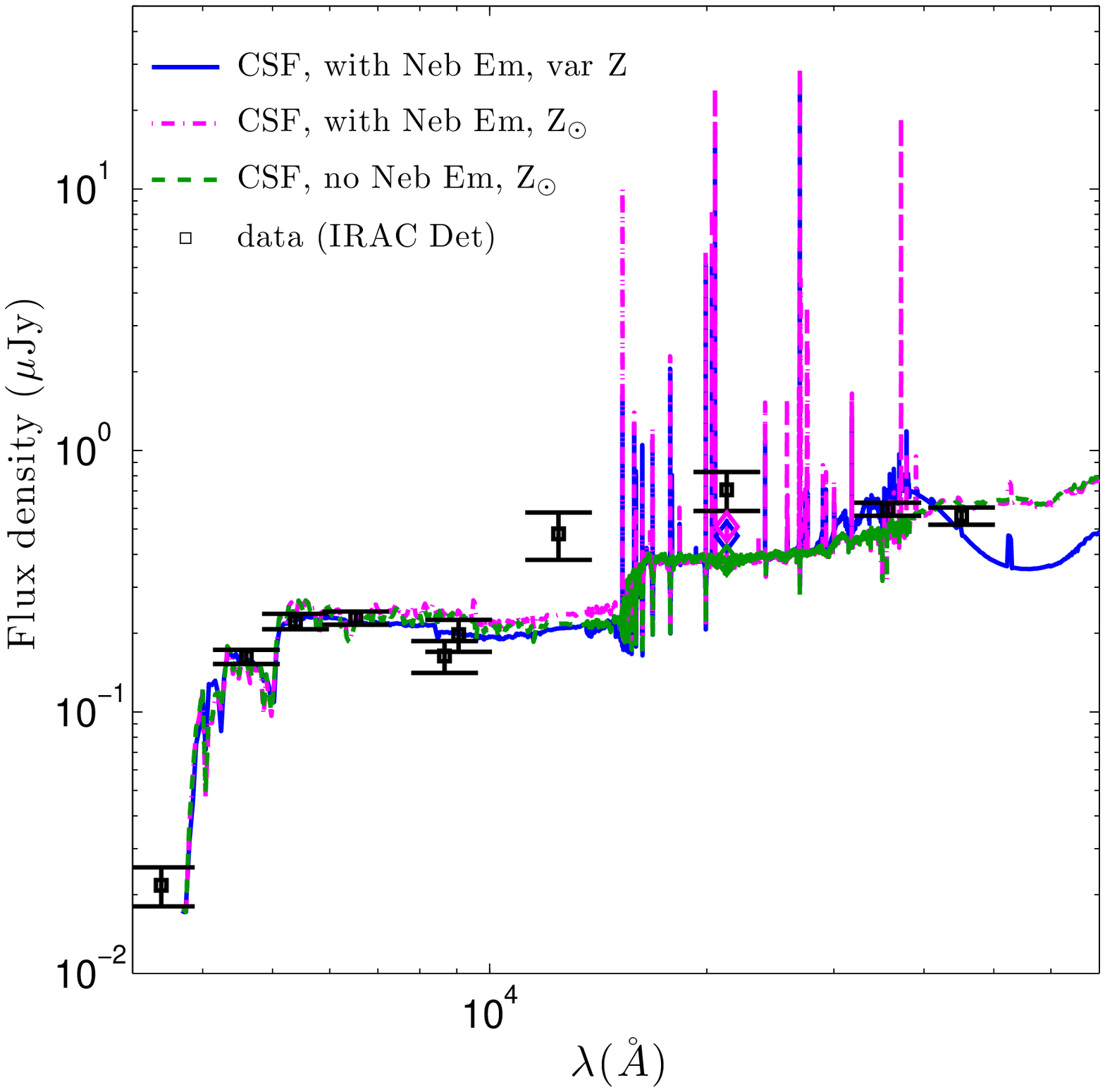}
\caption{Observed-frame data points (black) and best-fit SEDs for constant star formation history and different modeling assumptions presented in Table \ref{tab:syst}. The colored diamonds show the best fit point for CSF in the K-band, the most sensitive to nebular emission lines. {\bf Left:} IRAC Undetected sample. {\bf Right:} IRAC Detected sample. Data points with negative median stacked fluxes are not shown. The blue solid line shows the best-fit SED from the run with varying metallicity, corresponding to a value of $Z/Z_\odot$ = 0.005 for the Detected sample and $Z/Z_\odot$ = 0.006 for the Undetected sample.}
\label{fig:BF_SED}
\end{center}
\end{figure}

Adding the contribution of nebular emission produces little impact on the SED fit of the IRAC Detected sample. In fact, this population is older, and the relative strength of nebular emission lines and nebular continuum with respect to the stellar continuum is lower. Furthermore, the strongest emission lines (H$\alpha$, H$\beta$ and $O$III) mainly affect the K band; the corresponding data point was higher than the continuum SED, so that the SED parameters do not need to change to accommodate this feature, and instead the best-fit $\chi^2$ improves as a result. This behavior can be seen in the right-hand panel of Fig. 
\ref{fig:BF_SED}.
On the other hand, the inclusion of nebular emission has a strong effect both on the model SED and on the allowed parameter ranges for the younger LAEs in the IRAC Undetected sample, as seen in the left-hand panel of Fig. \ref{fig:BF_SED} and in Table \ref{tab:syst}. A new peak in the posterior probability distribution, at very young ages of $\sim$ 1 Myr and very low masses around 3 $\times 10^7$ M$_\odot$, is present. As a result, 
the stellar population is interpreted as significantly younger and less massive when nebular emission is taken into account.
This response of high-redshift galaxies to SED fitting including nebular emission is in agreement with the findings of \cite{2009A&A...502..423S}. 

 Finally, we consider the effect of using the BC03 stellar templates rather then the newer BC10 ones. Because of the differences in the templates discussed in Sec. \ref{sec:sps}, when BC03 models are used, older ages and higher stellar masses are needed to fit the SEDs of both samples.

The last set of assumptions in SPS modeling (no nebular emission, solar metallicity, BC03 templates) coincides with the previous analysis of the same samples presented in L08.
This allows a direct comparison of the best-fit approach vs the use of mean values computed from the posterior distribution, shown in Fig. \ref{fig:Systematics_All} and in Table \ref{tab:syst}. To compare the values of stellar masses on the same basis, we need to account for the mass loss due to the life cycle of stars, which we consider and was not explicitly quoted in that paper. We do so by multiplying the instantaneous star formation rate by the age reported in L08 to obtain the total mass transformed into stars M$_{\rm tot}$, and transform it into stellar mass using the conversion factor output by our code for the same model.   
Our best-fit parameters (also shown in the figure as magenta dot-dashed vertical lines) are in exact agreement with those of L08;
 however, the best-fit values do not lie exactly at the mean of the marginalized probability distribution, leading to a moderate shift of ages and stellar masses. The disagreement is within the $68\%$ confidence level and is expected when the N-dimensional posterior distribution is asymmetric.

\section{Conclusions}
\label{sec:conc}
We built a Markov Chain Monte Carlo code for SED fitting, GalMC, designed to determine physical properties of galaxies including age, stellar mass, dust content, metallicity, star formation history and redshift (the latter was fixed in the present analysis). The purpose of MCMC codes is to recover the probability distribution function (PDF) of the fit parameters by building a number of independent \textit{samples} with the same statistical properties as the PDF. In the Bayesian approach to statistical inference the relevant probability distribution is the \textit{posterior} probability, which depends on the likelihood of the parameters given the underlying data and on our prior beliefs about the model parameters.
Samples are built by exploring the parameter space in such a way that the density of sampled points 
is proportional to the probability distribution that we want to map. 
Once the samples have been obtained, they can be used to easily calculate the expectation values (\ie means) of parameters and the associated uncertainties at any confidence level, since integrals in any dimensions become tractable sums. Because most of the CPU time is spent in high-likelihood, informative regions of parameter space, MCMC algorithms offer a substantial improvement in efficiency with respect to methods based on mapping the posterior probability on a grid of reference parameter values. We found that this speed-up factor is of order $\sim$ 100 for a three-dimensional parameter space, and would rapidly increase for a larger number of parameters, since the expected scaling is linear in the number of parameters for MCMC, and exponential for grid-based methods. MCMC codes also offer warning flags of unreliable results by means of convergence tests 
and allow one to choose the sampling step size adaptively in each direction. 

We conducted an extensive series of tests on Mock catalogs built from a variety of stellar populations to ensure that the input parameters and their uncertainties were correctly recovered by GalMC, and to investigate the effect of using different priors. We then used the code to
determine the physical properties of two samples of LAEs at redshift z = 3.1. Our analysis showed that, on average, the LAEs in the IRAC detected sample are older and more massive than their counterparts in the IRAC undetected sample, and that both populations are essentially dust-free.  Furthermore, LAEs at $z$ = 3.1 have metallicities significantly lower than Solar, in agreement with recent spectroscopic studies of high-redshift LAEs. We investigated different assumptions on the star formation history of the LAEs, performing the fit with either constant or exponential (including increasing and decreasing) star formation rate, and found that the constant star formation model is favored by the data. Our results essentially confirm the findings of L08 about the distinct nature of the IRAC Detected and Undetected populations. However, important differences are found with respect to the previous analysis. The use of the improved SPS models from Charlot and Bruzual 2010 (as opposed to those from 2003) shifts the age of the older IRAC-detected population toward younger values, and therefore the mass toward lower values. Further changes are caused by our use of varying metallicity, by the inclusion of the flux from nebular continuum and emission lines, and by the use of the expectation values of parameters computed from the posterior distribution rather than best-fit values. The quoted uncertainties also differ, since we use the Bayesian approach and compute them as the 68\% quantile of the marginalized posterior distribution; for the $i$-th parameter, this means that we integrate the posterior distribution in all but the $i-$th direction. 

The algorithm development and SED analysis conducted in the present work set the foundation for a range of future applications. 
Our analysis showed that not only the parameter expectation values, but also their uncertainties, depend on the assumptions made in the SPS modeling; for example, the inclusion of nebular emission noticeably worsens the ability to rule out young ages for the IRAC Undetected sample. 
In the present work we have compared two models with differing number of parameters (a constant and an exponential star formation one). For the LAEs samples we studied, the MCMC sampling was directed toward very large values of the parameter $\tau$, similar to the CSF value, even when ESF was used, and the improvement in the quality of the fit due to use of one additional parameter was modest. We could conclude that the data do not strongly prefer ESF to CSF models. However, to better quantify this statement, or in general to assess whether the data favor the inclusion of additional parameters, we plan to use model selection \citep{citeulike:5645930}.

SED fitting is a technique of ever-increasing importance in astronomy, and it is essential that the tools used for statistical analysis keep up with the fast pace of the improvement in the data. Large-volume photometric surveys allow us to explore new and exciting directions, and this must be done while avoiding biasing assumptions on the shape of the probability function, and while maximizing the accuracy in the reported uncertainties. Markov Chain Monte Carlo algorithms enable efficient, reliable estimation of parameter expectation values and uncertainties, are suitable for exploring parameter spaces of high dimensionality, and are able to reveal degeneracies among parameters. GalMC is our implementation of this approach, and we hope it will prove useful for a wide range of astrophysical applications.

\acknowledgments

It is a pleasure to thank Amir Hajian, Ross Fadely and Chuck Keeton for many useful conversations on MCMC codes and shell scripts, Antony Lewis and Sarah Bridle for making CosmoMC publicly available and for helpful advice, Gustavo Bruzual and Stephane Charlot for providing and commenting the newest GALAXEV code and templates, Daniel Schaerer for help on the inclusion of nebular emission, Kamson Lai for useful conversations on the photometry of the LAEs, Mike Berry, Nick Bond, Robin Ciardullo, Niv Drory, Harold Francke, Caryl Gronwall, Peter Kurczynski, Kim Nilsson, Nelson Padilla, Kevin Schawinski, Licia Verde, Jean Walker, and Grant Wilson for helpful comments during the completion of this project and on the manuscript draft. We thank the anonymous referee for many expert comments and suggestions that helped us to improve the paper significantly.
This material is based on work supported by NASA through an award issued by JPL/Caltech and by 
the National Science Foundation under grant AST-0807570. Partial financial support for V.A. was provided by the National Science
Foundation under the PIRE program (award number OISE-0530095).
E.G. thanks the Department of Physics at U.C. Davis for hospitality during the completion of this research.

\newpage

\newpage

\clearpage

\begin{table}[h*]
  \center
  {\footnotesize
\begin{tabular}{@{\extracolsep{\fill}}lcccc}
    \hline
    \hline
Parameter & Symbol & Type $\quad $  & $\quad $Prior $\quad $ &  Range \\
\hline
Age since star formation begun& Age & Input & flat in $\log$(Age) & $\log_{10}$(Age/yr) $\in$ [6.0-9.33] \\ 
\hline
Total mass processed into stars & M$_{\rm tot}$ & Input & flat in $\log$(M$_{\rm tot}$) & $\log_{10}$(M$_{\rm tot}$/M$_\odot$) $\in$ [4.0-15.0] \\
\hline
Dust reddening  & E(B-V) & Input & flat in E(B-V) &$\quad $ E(B-V) $\in$ [0.0-1.0]\\
\hline
E-folding time   & $\tau$  & Input & flat in $\log(\tau)$ & $\quad $ $\tau \in$ [-4 Gyr, 4 Gyr]\\
\hline
Metallicity & $Z/Z_\odot$  & Input & flat in $\log(Z/Z_\odot)$ & $\quad $ $\log_{10}(Z/Z_\odot) \in$ [-2.3, 0.7]\\
\hline
Stellar Mass at t = Age  & M$_{*}$ & Derived  &$\quad $  \\
\hline
\hline
  \end{tabular}
  }
\caption{Input parameters of the MCMC code, priors used in the SED fit analysis, and derived parameters used to report results. The e-folding time $\tau$ is always the input variable of the SPS code, but we have chosen use $1/\tau$ as the MCMC sampling variable, as discussed in the text.}
\label{tab:param} 
\end{table}

% \clearpage

\begin{table}[h!]
  \center
  {\footnotesize
  \resizebox{\textwidth}{!}{%
\begin{tabular}{lccccccc}
    \hline
    \hline
    Sample & SFH & Z/Z$_{\odot}$ & Age (Gyr) & E(B-V) & M$^*$ ($10^8$ M/M$_\odot$)& $\tau$ (Gyr) & best fit $\chi^2$/d.o.f. \\ 
    \hline
    \hline
 IRAC Det & CSF & 0.02 & $0.67 [0.38 - 1.2] $ & $ 0.038 [0 - 0.047]$ & $31 [24 - 39] $ & $ \infty $ & 13.4/9 \\ 
    \hline
 	& ESF & 0.02 & $0.83 [0.54-2.1] $ & $ 0.046 [0-0.059] $ & $ 29 [23-38] $ & $ -3.0 [|\tau| > 0.67]$ & 10.8/8 \\
    \hline			       
    \hline
    IRAC Und    & CSF & 0.02 & 0.05 [0.04 - 0.16]  & 0.049 [0 - 0.064] & 1.7 [0.6- 3.0]& $\infty$ & 7.6/9\\
    \hline
    		       & ESF & 0.02 & 0.22 [0.001 - 2.1]& 0.047 [0 - 0.059] & 2.0 [1.2 - 3.1] & -2.7 $[|\tau| > 0.12]$ &  5.2/8 \\
    \hline
     \hline
  \end{tabular}
  }}
  \vspace*{0.3cm}
\caption{Results of SED fitting for constant and exponential star formation history. We report mean expectation values and 68\% confidence regions of parameters. For exponential SFH the parameter $\tau$ is close to the CSF value (defined by $|\tau|$ = 4 Gyr because of the priors used in ESF models). For both samples there is a modest improvement in the best-fit $\chi^2$ when using exponential SFH, but CSF is included in the range of values allowed at $68\%$ confidence, showing that the data do not favor ESF significantly.}
\label{tab:results} 
\end{table}

% \clearpage

\begin{table}[h!]
  \center
  {\footnotesize
  \resizebox{\textwidth}{!}{%
\begin{tabular}{lccccccc}
    \hline
    \hline
    Sample & Neb & SPS & Z/Z$_{\odot}$ & Age (Gyr) & E(B-V) & M$^*$ ($10^8$ M/M$_\odot$)& best fit $\chi^2$/d.o f. \\ 
    \hline
    \hline
    IRAC Det & Y & CB10 & 0.036[0.005 - 0.07] & $0.67 [0.37 - 1.18] $ & $ 0.046 [0 - 0.05]$ & $32 [25 - 42] $ & 13.2/8 \\ 
    \hline
  & Y & CB10 & 1 & $ 0.47[0.24 - 0.94] $ & $ 0.039 [0 - 0.041]$ & 29 [21 - 42]  & 18.6/9 \\ 
  \hline
& N & CB10 & 1 & $0.72 [0.33 - 1.3] $ & $ 0.028 [0 - 0.035]$ & $36 [28 - 47] $ & 19.6/9 \\ 
\hline
& N & BC03 & 1 & $ 1.0 [0.69 - 1.6] $ & $ 0.025 [0 - 0.03] $ & $ 55 [43 - 71] $ & 15.5/9 \\ 
\hline 
L08 Det &  N & BC03 & 1 & 1.6 [1.2 - 2.0] & [0 - 0.1] & 64 [43 - 85] & -- \\
    \hline			       
    \hline
     IRAC Und    & Y &  CB10 & 0.05 [0.005 - 0.13] & 0.06 [0.01 - 0.2]  & 0.066 [0 - 0.085] & 2.0 [1.1- 3.4]& 6.84/8\\
    \hline
    		     & Y &  CB10 & 1 & 0.007 [0.001 - 0.018]  & 0.035 [0 - 0.045] & 0.47 [0.26 - 0.98] & 9.92/9\\
    \hline
  		  & N &  CB10 & 1 & 0.049 [0.021 - 0.12]  & 0.04 [0 - 0.056] & 1.4 [0.89- 2.4] & 11.7/9\\
		  \hline
		  
		  & N & BC03 & 1 & 0.052 [0.02 - 0.13] & 0.039 [0 - 0.049] & 1.5 [0.92 - 2.7] & 11.8/9 \\
     \hline
     L08 Und & N & BC03 & 1 & 0.16 [0.05 - 0.27] & 0 [0 - 0.1] & 2.6 [0.83 - 5.7] & -- \\
     \hline
     \hline
  \end{tabular}}
  }
\caption{Mean expectation values and 68\% confidence regions from SED fitting for constant star formation history, using different assumptions on inclusion of nebular emission, SPS modeling and metallicity. Solar metallicity is excluded at 95$\%$ confidence by the data; see text for more details. For comparison, we also report best-fit values and 68\% confidence regions obtained for the same samples by L08.}
\label{tab:syst} 
\end{table}

\end{document}